\input epsf
\documentstyle[12pt]{article}
\begin{document}
\begin{titlepage}
\setcounter{page}{0} 

\begin{center}
{\Large   NLL Corrections for B-Meson  Radiative Exclusive Decays}\\
\vspace{2cm}
{\large H. H. Asatryan$^a$, H. M. Asatrian$^a$, D. Wyler$^b$}\\
\vspace*{0.5cm}

\centerline{{\rm ${}^a$}Yerevan Physics Institute,
Alikhanyan Br., 375036-Yerevan, Armenia}
\centerline{{\rm ${}^b$} Institute for Theoretical Physics, University 
of Z\"urich, Z\"urich, Switzerland}

\vspace{3cm}
\end{center}
\centerline{{\bf{Abstract}}}
\vspace{0.5cm}
We calculate the next-to-leading corrections to the
branching ratio of exclusive $B\to K^* \gamma$ decay. 
The renormalization scale dependence is reduced compared to the 
leading logarithmic result but there remains a dependence on a cutoff 
parameter of the hadronic model. The calculated corrections increase
the predicted branching ratio by about 10\%, but it remains in agreement 
with the experimental value.

\end{titlepage}
\newpage
\section{Introduction}
The rare neutral flavor changing decays of B-meson are
generally considered a good testing ground for new physics \cite{joan}. 
The inclusive decay $b \to s \gamma$ has therefore received 
intensive experimental \cite{AMMARCLEO} and theoretical 
attention \cite{GSWISE}-\cite{CDGG}. In particular, the 
next-to-leading order calculation for the standard model
strongly improved the theoretical uncertainties, most notably 
the scale dependence of the perturbative calculation.

The calculation of the decay rates for the
corresponding exclusive decays such
as $B\to K^* \gamma$ \cite{ALAMCLEO} is usually much more
complex because of bound state effects (for a recent work, see for
instance \cite{wise}). But they are easier to
observe, and so it is also of interest to calculate their
branching ratios and direct CP asymmetries to the highest 
possible precision. In this paper we consider the branching ratio 
for $B\to K^* \gamma$ in the next-to-leading logarithmic approximation.
The decay $B\to \rho \gamma$ can in principle be treated in the
same way, but there are several additional contributions which
require new methods.  

In \cite{GSW}, the $B\to \rho \gamma$ and
$B\to K^*\gamma$ branching ratios and rate asymmetries
were investigated in the
leading logarithmic approximation within a simple meson model. Although
this model does not give completely correct results (in particular the
ratio of the semileptonic $\pi$ and $\rho$ decays),
we will also use it in this work, because of its simplicity.
 
As usual, the basis of the calculation is the
effective Hamiltonian which we record here
for completeness. For  $B\to K^* \gamma$ decay it has the
form \cite{GSW}
\begin{equation}
{\cal H}_{eff}=-\frac{4G_F}{\sqrt{2}}
\left \{\Sigma_iv_tC_i(\mu)O_i(\mu)-                     
v_u\left [C_1(\mu)(O_{1u}(\mu)-O_1(\mu))+
C_2(\mu)(O_{{2u}}(\mu)-O_2(\mu))\right ]\right \},
\end{equation}
where $v_a=V_{tb}V_{as}^*$, a=u,c,t, i=1,2,...8. For the decay
$B\to \rho \gamma$ one must replace $v_a$ by $v_a^{'}$,
with  $v_a^{'}=V_{tb}V_{ad}^*$.
The operators $O_i$ can be found in many places \cite{GHW,AAG,BURAS}.

\section{The bound state model}

As mentioned, we will use a simple model for the mesons to take
into account the bound state effects \cite{GSW,GW} (for other approaches,
see \cite{other},\cite{ABS}).
The  B and the $K^*$ are described by two constituents:
$B=(b,\bar{q})$, $K^*=(s,\bar{q})$. 
The sum of the four momenta of the constituents
equals that of the bound state. The mass of the spectator
antiquarks is taken to be zero and the b-quark mass becomes
momentum dependent:
\begin{equation}
m_b=\sqrt{m_B^2-2p_Bp_q},                      
\end{equation}
where $p_B$ and $p_q$ are the four momenta
of the B-meson and spectator quark, respectively.
The momentum $p_q$ in the B-meson rest frame is restricted to
\begin{equation}
|p_q|=E<m_B/2.                                       
\end{equation}
The B meson is represented by a matrix $\Psi_B$
\begin{eqnarray}
\nonumber
&&\Psi_B=C_B\int \frac{d^3p_q}{2E_q(2\pi)^3}\sqrt{\frac{m_B}{2(m_B^2-p_Bp_q)}}
\phi_B(|p_Bp_q|/m_B)\Sigma_b \times \frac{1}{\sqrt{N}}1_N\\
&&\Sigma_B=-(\hat{p}_b+m_b)\gamma_5\hat{p}_q, \hspace{1cm}         
\phi_B(p)=\exp(-p^2/(2p_F^2)),
\end{eqnarray}
where $C_B$ is a normalization factor and $p_F$ is fixed in such a way
that one obtains the correct value of B-meson decay constant $f_B$. 
The 'hat' symbol
stands for the usual 'slash'. 

The final state vector meson $K^*$ is represented by two constituents
with parallel momenta $p_s=yp_V$, $p_q=(1-y)p_V$. 
In most cases its mass and 
that of the constituent $s$ quark can be neglected (see however the
discussion of that
point in section 4). It is described by the matrix
\begin{eqnarray}
\nonumber
&&\Psi_V=C_V\int_0^1 dy \phi_V(y)\Sigma_V\times \frac{1}{\sqrt{N}}1_N,
\hspace{1cm}
C_V=f_V/(4\sqrt{N})\\
&&\Sigma_V=\lim_{m_V\to 0} \hat{\epsilon}_V^*(\hat{p}_V+m_V),        
\hspace{1cm}
\phi_V(y)=6y(1-y)(1+\cdot \cdot \cdot),
\end{eqnarray}
where $\phi_V(y)$ for $V=K^*,\rho$ are given by \cite{ABS,EIM}:
\begin{eqnarray}
\nonumber
\phi_{K^*}(y)&=&6y(1-y)(1+0.57(2y-1)-1.35((2y-1)^2-1/5)+\\             
&+&0.46(7(2y-1)^3/3-(2y-1)))\\
\nonumber
\phi_{\rho}(y)&=&6y(1-y)(1-0.85((2y-1)^2-1/5)).
\end{eqnarray}
The $B\to K^*\gamma$ transition matrix elements have a following general form:
\begin{equation}
<V\gamma|O_i|B>=em_B\epsilon^{\mu}_{\gamma}\left [i\epsilon_{\mu \nu \alpha 
\beta}p_{\gamma}^{\nu}\epsilon_V^{\alpha}p_V^{\beta}F_1[O_i]+((p_{\gamma}\epsilon_V)p_{V\mu}-                          
(p_{\gamma}p_{V})\epsilon_{V\mu})F_5[O_i]\right ].
\end{equation}
For a massless final state meson we have $F_1=F_5$.

In our model, the hadronic matrix elements are written as
\begin{eqnarray}
\nonumber
<V\gamma|O_i|B> & = & C_BC_V\int\frac{d^3p_qdy}{(2\pi)^32E_q}\sqrt{\frac{m_B}
{2(m_B^2-p_Bp_q)}}\\                                               
&\times & \phi_B \left (\frac{p_Bp_q}{m_B}\right )\phi_V(y)\frac{1}{N}
Tr\left [\Sigma_VM_{sb}\Sigma_BM_{qq^{'}}\right ],
\end{eqnarray}
where the trace is in Dirac and color space and the matrices $M_{sb}$ and
$M_{qq^{'}}$ are related to the quark-level matrix elements by
\begin{equation}
<s\bar{q}^{'}\gamma|O_i|b\bar{q}>=\bar{u}_sM_{sb}u_b\bar{v}_q     
M_{qq^{'}}v_{q^{'}}.
\end{equation}
The decay width in NLL approximation can be written as
\begin{equation}
\Gamma (B\to V \gamma) = |v_t|^2G_F^2 \alpha_{em}m_B^5
\sum_l|\sum_iC_iF_l[O_i]|^2\left (1-\frac{8}{3}\frac{\alpha_s}{\pi}\right ).
\end{equation}
In leading approximation only $O_7$ contributes and the
last parenthesis is simply one.

In order to eliminate some of the uncertainties of the model, we
write
\begin{equation}
BR(B\to V\gamma)\equiv \frac{\Gamma(B\to V\gamma)}{\Gamma_{sl}}
BR(B\to X l\nu_l),                                              
\end{equation}
where $\Gamma_{sl}$ in NLL approximation is given by
\begin{eqnarray}
\nonumber
&&\Gamma_{sl}=\frac{G_F^2}{192\pi^3}|V_{cb}|^2C_B^2
\int \frac{d^3p_q}{(2\pi^3)}|\phi_B(|p_q|)|^2\frac{m_b^6}{m_B-E_q}g(r)
\left (1-\frac{2\alpha_s}{3\pi}f(r)\right )\\
&&g(r)=1-8r^2+8r^6-r^8-24r\log(r) \\                            
\nonumber
&&f(r)=(\pi^2-\frac{31}{4})(1-r)^2+\frac{3}{2}, \quad \quad r=m_c/m_b(p_q)
\end{eqnarray}
and we use the experimental value for
the semileptonic decay, $BR(B\to X l\nu_l)$=0.103 \cite{pdg}.

\section{The ``inclusive amplitudes''}

Consider first the decay graphs where the spectator is not
touched; they are very similar to the inclusive
decays. As mentioned, 
in leading logarithmic approximation only $F[O_7]$ 
contributes to $<V\gamma |{\cal H}|B>$ (Fig. 1). Evaluating 
the trace, one gets \cite{GSW}:
\begin{equation}
F^0[O_7]=-\frac{\sqrt{m_B}C_BC_V}{4\pi^2}\int dy                   
\frac{y}{\sqrt{1+y}}\phi_V(y)\phi_B \left (\frac{m_B(1-y))}{2}\right ).
\end{equation}
Here, and everywhere we will set $F_1=F_5 \equiv F$ for all form
factors. The deviation due to small final state masses can be
neglected.

For the NLL amplitudes, we can use the results for NLL inclusive
$O_7$, $O_2$, $O_2$ amplitudes given in \cite{GHW}
\begin{eqnarray}
\label{nll}
\nonumber
&&<s\gamma|O_{7}|b>=<s\gamma|O_7|b>_{tree}\left (1+\frac{\alpha_{s}(m_b)}{4\pi}
\left (\gamma^{(0)eff}_{77}\log(m_b/\mu)+r_7\right )\right )\\                       
\nonumber
&&<s\gamma|O_{2}|b>=<s\gamma|O_7|b>_{tree}\frac{\alpha_{s}(m_b)}{4\pi}
(l_2\log(m_b/\mu)+r_2)\\
&&<s\gamma|O_{8}|b>=<s\gamma|O_7|b>_{tree}\frac{\alpha_{s}(m_b)}{4\pi}
(l_8\log(m_b/\mu)+r_8)\\
\nonumber
&&<s\gamma|O_7|b>_{tree}=m_b\frac{e}{8\pi^2}\bar{u}(p^{'})
\hat{\epsilon}\hat{q}u(p), 
\end{eqnarray}
where the expressions for $r_2$, $r_7$, $r_8$, $l_2$, $\gamma^{(0)eff}_{77}$,
$l_8$ are
given in \cite{GHW}.\\

The problem of Bremsstrahlung corrections needs discussion.
In the inclusive case the infrared divergences from the
NLL calculation (loops)
of  $<s\gamma|O_{7}|b>$ (Fig. 2) cancels at the level of
the decay width when one adds Bremsstrahlung corrections
associated with the operator $O_7$
(that is when we sum over final states with additional soft gluons).
In the exclusive case the situation is much more complicated. Normally we
consider the mesons to contain only two constituents (quark and antiquark)
in a color singlet state. Therefore, it is not possible to simply add to
the Hilbert space of hadrons states with extra gluons; furthermore,
soft gluons are not physical. Instead, these
gluons must be incorporated into the physical hadrons, making it
necessary to consider three particle states (quark, antiquark and gluon)
etc for the $K^*$ meson and other physical particles.
For a discussion of this point see \cite{brod}.

This well known problem cannot be really solved with the present 
methods and one must resort to approximate methods
(Additional IR problems connected to the exchange of gluons
with the spectator quark will be discussed shortly).

We have therefore used two heuristic methods to obtain finite
results. In Method $1$ we use the fact that we could write 
the finite 'total inclusive amplitude' (including the bremsstrahlungsgluon) 
as an effective matrix element 
for $b \to s \gamma$ as in equation (3.5) of \cite{GHW}.
There, this was just a mathematical trick 
to obtain a convenient form of the result up to the 
desired accuracy. Physically, we might envisage it as a 
replacement of the eight gluons by an abelian vector. This 
interpretation follows if one calculates the cross section 
of the bremsstrahlungsprocess which consists
in summing over the final states and where the sum over the
gluons indeed yields an 'abelian' contribution (with the correct
factor). In this way the final result is simply

\begin{eqnarray}
\nonumber
F^{incl}[O_7] = &-&\frac{C_B C_V \sqrt{m_B}}{4\pi^2}
\frac{\alpha_{s}(m_b)}{4\pi}(\gamma^{(0)eff}_{77}\log(m_b/\mu)+r_7)
\times \\
\nonumber
&\times &\int dy \frac{y}{\sqrt(1+y)}\phi_V(y) \phi_B\left (\frac{m_B(1-y)}{2}
\right )\\                                                         
\nonumber
F^{incl}[O_8] = &-&\frac{C_B C_V \sqrt{m_B}}{4\pi^2}
\frac{\alpha_{s}(m_b)}{4\pi}(l_8\log(m_b/\mu)+r_8)
\times \\
&\times & \int dy \frac{y}{\sqrt(1+y)}
\phi_V(y) \phi_B\left (\frac{m_B(1-y)}{2}\right )\\
\nonumber
F^{incl}[O_2] = &-&\frac{C_B C_V \sqrt{m_B}}{4\pi^2}
\frac{\alpha_{s}(m_b)}{4\pi}(l_2\log(m_b/\mu)+r_2)
\times  \\
\nonumber
&\times &
\int dy \frac{y}{\sqrt(1+y)}
\phi_V(y) \phi_B\left (\frac{m_B(1-y)}{2}\right ).
\end{eqnarray}
In (3) and (4) $r_8$ and $r_2$ include real and imaginary part \cite{GHW}.\\

In Method $2$ we cut the virtual loop momenta at a
value $\Lambda$.
The same cutoff $\Lambda$ is then used for the momentum when
calculating the $O_7$ diagram (Fig. 3 a,b) with a gluon exchanged 
between the quark lines (next section). No Bremsstrahlung corrections 
are then added. 
The physical picture here is that gluons with momenta below 
$\Lambda$ are embodied in the hadronic wave function.  The formulae for 
the contribution of the operator $O_7$ with a cutoff are given 
in Appendix A.\\

\section{NLL ``exclusive'' amplitudes}

Next consider the 'exclusive' amplitudes, i.e. 
amplitudes which include the exchange of 
gluon between two quark lines. \\
As mentioned above, we require the gluons to be of-shell by at 
least an amount $\Lambda^2$. In the numerical evaluation, we will
consider the three cases $\Lambda$ =0.2,0.5,1GeV.\\

\subsection{Contributions from $O_2$}

The form factors associated with $O_2$ originate from
two types of diagrams (Fig. 4a, b).
When the photon is emitted from
external quark lines (Fig. 4a), 
we can use the expression for the effective $b\bar{s}g$ vertex
given in \cite{GHW,GSW}
\begin{eqnarray}
&&I^a_{\mu}=\frac{g_s}{16\pi^2}\left (q^2 \gamma_{\mu}-q_{\mu} \hat{q}\right )
L \frac{\lambda^a}{2}V\left (\frac{q^2}{m_i^2} \right ) \\
\nonumber
&&V\left (\frac{q^2}{m_i^2}\right )=4\Gamma(\epsilon)\mu^{2 \epsilon}
\exp(\gamma_E \epsilon)
(1-\epsilon)\exp(i\pi \epsilon)\int_0^1 dx
[x(1-x)]^{1-\epsilon} 
\left[ q^2-\frac{m_i^2}{x(1-x)}+i\delta \right]^{-\epsilon}.
\end{eqnarray}
The explicit calculation of V(r) $(r=\frac{q^2}{m_i^2})$ gives:
\begin{equation}
\label{vfactor}
V(r)=\frac{2}{3}\frac{1}{\epsilon}+\frac{2}{3}\log\frac{\mu^2}
{m_i^2}+\frac{4}{9}\frac{6+r}{r}+\frac{4}{3}                  
\frac{(r^2-2r-8)}{r\sqrt{r(4-r)}}\arctan \sqrt{\frac{r}{4-r}}
\end{equation}
for $r<4$, and
\begin{eqnarray}
\nonumber
&&V(r)=\frac{2}{3}\frac{1}{\epsilon}+\frac{2}{3}\log\frac{\mu^2}{m_i^2}+
\frac{4}{9}\frac{6+r}{r}+\frac{2}{3}
\frac{(r^2-2r-8)}{r\sqrt{r(4-r)}}\log\frac{\sqrt{r}-\sqrt{r-4}}   
{ \sqrt{r}+\sqrt{r-4}}+\\
&&+\frac{2}{3}i\frac{r+2}{r}\sqrt{\frac{r-4}{r}}
\end{eqnarray}
for $r>4$. We see that $V(r)$ contains an infinity $\frac{1}{\epsilon}$,
where $\epsilon$ is the parameter which comes from the dimensional
regularization method used. It is of course cancelled by an
appropriate counter-term. The contribution from $V(r)$ 
to the effective Hamiltonian then has the following form
\begin{equation}
C_2V(r)\frac{\alpha_s}{4\pi}\frac{1}{q^2}\left (\bar{s}\frac{1}{2}\lambda^a
(\gamma_{\mu} q^2-q_{\mu}\hat{q})Lb\right )\left (\bar{u}\gamma_{\mu}\frac{1}{2}  
\lambda^a u \right ),
\end{equation}
while that of the counter-term reads:
\begin{eqnarray}
\nonumber
&&-\frac{4}{3}C_2\frac{\alpha_s}
{48\pi \epsilon}<s\bar{u}\gamma|O_3|b\bar{u}>                   
+3\frac{4}{3}C_2\frac{\alpha_s}{48\pi \epsilon}
<s\bar{u}\gamma|O_4|b\bar{u}>-\\
&&-\frac{4}{3}C_2\frac{\alpha_s}{48\pi \epsilon}
<s\bar{u}\gamma|O_5|b\bar{u}>
+3\frac{4}{3}C_2\frac{\alpha_s}{48\pi \epsilon}                
<s\bar{u}\gamma|O_6|b\bar{u}>.
\end{eqnarray}
It is easy to see the cancellation of the divergences proportional to
$\frac{1}{\epsilon}$ using the formula
$\sum_r\frac{1}{2}\lambda^r_{ab}\frac{1}{2}\lambda^r_{cd}=
\frac{1}{6}(3\delta_{ad}\delta_{bc}-\delta_{ab}\delta_{cd})$.\\
Summing over all diagrams
where the photon is emitted from different external quark lines we 
finally obtain for the form-factor in the rest frame of the B-meson
\begin{eqnarray}
\nonumber
F^1[O_2]&=&\frac{g_{s}^2}{16\pi^2}\frac{N^2-1}{2N}\frac{1}{m_B^2}
\int \phi_B(E_q)\phi_V(y)\frac{1}{\sqrt{2(m_B-E_q)}}\frac{E_{q}^2}{\pi^2}
dE_{q}dzdy\times\\                                                      
&\times&\left (-Q_d\frac{V(r_1)}{y}+Q_u\frac{V(r_2)}{1-y}\right ),
\end{eqnarray}
where $Q_u=2/3$, $Q_d=-1/3$, $r_1=((1-y)p_V-p_q)^2$, 
$r_2=((1-y)p_V+p_\gamma-p_q)^2$, $z=\cos\theta$ and $\theta$ is the
angle between the spectator quarks in the $B$ and  
$K^*$-mesons, respectively.\\
When the photon is emitted from the quarks in the
internal loop (Fig 4b), we will 
use the following expression for the effective
$\bar{s}bg\gamma$-vertex \cite{GSW}:
\begin{eqnarray}
\nonumber
&&I^a_{\mu\nu}=-\frac{g_{s}eQ_u}{8\pi^2}\frac{\lambda^a}{2}
\left [i\epsilon_{\beta\mu\nu\alpha}(q^{\beta}\Delta i_5              
+p_{\gamma}^{\beta}\Delta i_6)+ \right. \\
&&\left.+i\frac{\epsilon_{\rho\sigma\mu\alpha}}{p_{\gamma}\cdot q}q^{\rho}
p_{\gamma}^\sigma
q_{\nu}\Delta 
i_{23}+i\frac{\epsilon_{\rho\sigma\nu\alpha}}{p_{\gamma}\cdot q}
q^{\rho}p_{\gamma}^{\sigma}p_{\gamma \mu}\Delta i_{26}
\right ]\gamma^{\alpha}L,
\end{eqnarray}
where q is the momentum of the gluon. The quantities $\Delta i_{n}=
\Delta i_{n}(z_{0},z_{1},z_{2})$ are functions of the variables
$z_{0}=s/m_{i}^{2},z_1=q^2/m_i^2,z_2=p_{\gamma}^2/m_i^2$, and s is
the invariant mass of the internal quark pair. When the
photon is on-shell we have
\begin{eqnarray}
\nonumber
\Delta i_5(z_0,z_1,0)=-1+\frac{z_1}{z_0-z_1}\left (Q_0(z_0)-Q_0(z_1)\right )-
\frac{2}{z_0-z_1}\left (Q_-(z_0)-Q_-(z_1)\right )\\
\Delta i_6(z_0,z_1,0)=1+\frac{z_1}{z_0-z_1}
\left (Q_0(z_0)-Q_0(z_1)\right )+   
\frac{2}{z_0-z_1}\left (Q_-(z_0)-Q_-(z_1)\right )\\
\nonumber
\Delta i_{23}(z_0,z_1,0)=\Delta i_5(z_0,z_1,0)=-\Delta i_{26}(z_0,z_1,0).
\end{eqnarray}
The functions $Q_0$ and $Q_-$ are defined in the following way:
\begin{eqnarray}
Q_-(x)=-2\pi i\log\left (\frac{\sqrt{x}+\sqrt{x-4}}{2}\right )-              
x\int_{0}^{1}du\frac{(2u-1)\log u}{1-xu(1-u)}\\
Q_0(x)=-\pi i\sqrt{\frac{x-4}{x}}-
x\int_0^1du\frac{u(2u-1)}{1-xu(1-u)}.                           
\end{eqnarray}

The expression for the form factor is the following:
\begin{eqnarray}
&&F^2[O_2]=\alpha_s C_BC_VC_N                                  
\int \phi_B(E_q)\phi_V(y)\frac{Q_ue^2}
{\sqrt{2(m_B-E_q)}}\frac{E_{q}^2}{16q^2\pi^3}
dE_{q}dzdy \times\\
\nonumber
&&\times\left (\frac{4(1-z)E_q}{m_B}\Delta_{i_{5}}
-4\Delta_{i_6}+2\frac{(1-z^2)E_q^2}
{p_{\gamma}\cdot q}\Delta_{i_{23}}+
\frac{(m_b(1-y)-2E_q)(1+z)m_b}{p_{\gamma}\cdot q}\Delta_{i_{26}}\right ).
\end{eqnarray}

\subsection{Contributions from $O_8$ and $O_7$}

The contribution of the diagrams where the photon is emitted
from the 
spectator quark lines (Fig. 5) in the $B$ and $K$ mesons equals
\begin{eqnarray}
\nonumber
F^1[O_8] & = & C_B C_V C_N Q_u\frac{\alpha_s}{4\pi^3}\frac{1}{m_B^2}
\int dE_q dzdy\frac{E_q(m_B-2E_q)}{(1-y)\sqrt{2(m_B-E_q)}} \\
&\times &\frac{E_q(2-y(1-z))}{m_B(1-y)-E_q(2-y)(1-z)}            
\phi_B(E_q)\phi_v(y).
\end{eqnarray}
Similarly, when the photon radiates
from b- and s-quark lines (Fig. 5), one obtains
\begin{eqnarray}
F^2[O_8]=C_B C_V C_N Q_d\frac{\alpha_s}{4\pi^3}\frac{1}{m_B^2}
\int dE_q dzdy\frac{E_q(m_B-2E_q)}{y\sqrt{2(m_B-E_q)}}         
\phi_B(E_q)\phi_v(y).
\end{eqnarray}

The contribution associated with the operator
$O_7$ (Fig. 3 a,b) has been calculated in \cite{GSW}:
\begin{eqnarray}
\nonumber
&&F[O_{7}]=\frac{g_{s}^{2}m_{B}}{16\pi^{4}}C_{N}C_{B}C_{V}
\int dydE_qdz\phi_{V}(y)\phi_{B}(E_{q})                        
\frac{E_q^{2}(m_{B}-2E_{q})(2-y(1-z))}
{N_{b}q^{2}\sqrt{2(m_{B}-E_{q})}},\\
&&q^2=\left [(1-y)m_{K^*}^2-\frac{E_q}{m_B}\left (m_B^2(1-z)+
m_{K^*}^2(1+z)\right ) \right ](1-y)\\
\nonumber 
&&N_b=2E_qm_B-m_B^2(1-y)-m_{K^*}^2y(1-y).     
\end{eqnarray}

In general, we neglect the mass of the $K^*$ meson and of its
constituents. 
However, when we deal with infrared divergent 
integrals and use the cutoff parameter $\Lambda$, we cannot neglect 
$m_{K^*}$ if it is larger than $ \Lambda$. 
Therefore we take into account the
finite values of $m_{K^*}$ and $m_{s}$ when calculating the matrix elements of
$O_7$ (Figs. 2 and 3).

\subsection{Exchange diagrams}

Exchange diagrams, which can contribute to the decays
$B\to V \gamma$
are represented in Fig. 6 for the
operators $O_{1u}$, $O_{2u}$,
$O_3$, $O_4$, $O_5$ and $O_6$.\\
For $B\to K^*\gamma$ decay the contributions of the operators
$O_{1u}$, $O_{2u}$ are proportional to the small CKM factor
$V_{ub}V_{us}^*$ and consequently can be neglected.
Also the contributions of the operators $O_3$, $O_4$
can be neglected in the approximation where the vector meson mass
is equal to zero.
The contribution from the operator $O_5$ comes from the
exchange diagrams shown in
Fig. 6. The expression for corresponding form factor is the following:
\begin{equation}
F[O_5]=-\frac{2}{3\pi^2}\frac{C_{B}C_{V}}{m_{B}^2}\int E_q^2\frac{\phi_B(E_q)
\phi_V(y)}{\sqrt{2(m_B-E_q)}}\frac{(1+y)}{y(1-y)}dE_qdy.        
\end{equation}
The form factor for $O_6$ is given by the same formula except for an
additional factor 3 coming from the trace in color space:
$F[O_6]=3F[O_5]$.\\

\subsection{Renormalization scale dependence}

It is known that NLL contributions drastically reduce the large $\mu$
dependence of inclusive decay rate for $B\to X_{s}\gamma$;
more exactly, the $\mu$ dependence of the LL term proportional to
$C_7^2$ is canceled by explicit logarithms, proportional to $\alpha_s$.  
As we have here an additional $\mu$ dependence in $F^1[O_2]$ and also 
in $C_5$, $C_6$ in the exchange 
diagrams, it is necessary to check whether such 
a cancellation also materializes in the
exclusive decays $B\to V\gamma$.
The $\mu$ dependence
of the contributions of the
operators $O_5$, $O_6$ is due to the $\mu$ dependence
of the Wilson coefficients $C_5$, $C_6$ which is
determined 
by the renormalization group equation:
\begin{equation}
\frac{dC_i}{d(\log\mu)}=\frac{\alpha_s}{4\pi}\gamma_{ji}^0C_j   
\end{equation}
An approximative solution for $C_5$, $C_6$ is
\begin{displaymath}
C_5=\frac{\alpha_s}{4\pi}
\gamma_{25}^0\log(\mu/M_W)
\quad \quad \quad
C_6=\frac{\alpha_s}{4\pi}
\gamma_{26}^0\log(\mu/M_W),
\end{displaymath}
where $\gamma_{25}^0$=-2/9, $\gamma_{26}^0$=2/3.
The exchange diagrams contribute in the combination $(3C_6+C_5)F[O_5]$. 
On the other hand the scale dependence originating from
$F[O_2]$ is the same and opposite as seen from Eq. (\ref{vfactor}) and
the color matrix identity given there.

\section{Numerical results and discussion}

The decay amplitude is given by the following expression:
\begin{eqnarray}
\nonumber
{\cal M}(B\to K^* \gamma)&=&\frac{4G_F}{\sqrt{2}}
em_B\epsilon^{\mu}_{\gamma}v_t
\left (i\epsilon_{\mu \nu \alpha 
\beta}p_{\gamma}^{\nu}\epsilon_V^{\alpha}p_V^{\beta}+
(p_{\gamma}\epsilon_V)p_{V\mu}-         
(p_{\gamma}p_{V})\epsilon_{V\mu}\right )
\\
&\times & \left  ( C_7F^0[O_7]
+C_7F^{incl}[O_7]+C_2F^{incl}[O_2]+C_8F^{incl}[O_8] \right.\\
\nonumber
& +& \left. C_2F[O_2]+C_8F[O_8]+C_7F[O_7]+(C_5+3C_6)F[O_5]\right ),
\end{eqnarray}
with $F[O_2]=F^1[O_2]+F^2[O_2]$,
$F[O_8]=F^1[O_8]+F^2[O_8]$.\\

We begin with a discussion of the individual contributions for one
special case: method 2 and $\Lambda =0.2$, $\mu =5GeV$, $f_B=160 $MeV 
\cite{AOKI}. The contribution associated with the operator $O_7$  consists of
two pieces: an "inclusive" part (Fig. 2) and an "exclusive" part (Fig. 3). 
The contribution of the first diagram (to the total rate) 
is -37\%, while that of 
the second is -20\% 
\footnote{Since we only calculate to order $\alpha_s$, all
corrections we calculate contribute linearly.}. 
Similarly, the contribution of the operator $O_2$ 
is altogether 55\%. 51\% come from the  "inclusive" 
diagrams (Figs. 3,4 of \cite{GHW}), while the 
exclusive contributions amount to 5\% (Fig. 4a) and -1 (Fig. 4b).
The operator $O_8$ contributes only 2.3\%, which 
consist of 3.8\% for the
"inclusive" part (Fig. 9 of 
\cite{GHW}) and of -1.5\% for the "exclusive" part (Fig. 5).
The contribution of the diagrams associated with the operators 
$O_5$ and $O_6$ is 6.2\%.  The $O(\alpha_s)$ corrections to $C_7$ 
amount to -5\%. The contributions of the additional 
factors in (10) and (12) equal -7\%.\\

Our numerical results for the $B \to K^* \gamma$ decay rate are given in
Table 1 in LL and NLL approximation (for the two methods to regularize the IR
divergences). The dependence on the cutoff
$\Lambda$ is shown in Fig. 7, while the variation with 
$f_B$ is given in Fig. 8. For $f_B=160MeV$ and $\Lambda$=0.5GeV 
our prediction for the branching ratio is $(5.18\pm 0.5) \cdot 10^{-5}$ 
(method 1) and $(5.16\pm 0.5) \cdot 10^{-5}$ (method 2)
(the errors refer to the scale uncertainty). 
For both methods, they are in agreement with the experimental
value $BR(B\to K^* \gamma)=(4.2\pm 0.8 \pm 0.6 )\times 10^{-5}$ 
\cite{AMMARCLEO,SKW}, although it is somewhat on the high side.
We note that with the simple expression for the vector
meson wave function $\phi_V(y)=6y(1-y)$, the prediction for the branching
is lowered by $\sim 15\%$.  
The NLL contributions clearly reduce 
the scale dependence of the LL result. For the decay rate in LL approximation, 
the uncertainty due to the scale dependence is 27\%. On the other
hand, the NLL results vary between 4\% and 11\% for method 1 and
between 4\% and 21\% for method 2 for $1GeV>\Lambda >0.2GeV$.
This remaining scale dependence is larger than the one of
the inclusive decay. However, the standard model 
yields indeed an 'unnaturally' low scale dependence for $b \to s \gamma$
\cite{neub}. Therefore the variation with the scale seems reasonable
to us. 

The uncertainty due to the unknown value of $\Lambda$ is 
between 6\% and
14\% for method 1 and between 10\% and 26\% for method 2.
The results for the two methods for $\mu=5GeV$ and $0.2GeV<\Lambda<1GeV$ 
differ no more  than 14\%. Thus
in method 1, the theoretical uncertainty is quite small, in the vicinity
of 20\%. If this is a correct assessment, then it is clearly
desirable to improve the experimental accuracy.

The rather strong $\Lambda$ dependence of method 2 is clearly a 
shortcoming of this work. It arises mainly from the logarithmic
cutoff singularity of the gluonic corrections to the operator $O_7$.  
A clear improvement would be to take into account
three body Fock states (with an extra gluon) whose matrix
element could remove the $\Lambda$ divergence. This requires
relations between two-and three-body wave-functions. In simple
cases, such relations exist \cite{anto}. However, a realistic calculation
is beyond the state-of-the art.
Another source of $\Lambda$ dependence lies in the "inclusive"
contributions associated with the operators $O_2$ and $O_8$. To
isolate it, one would need a two-loop calculation with a finite cutoff; 
this is beyond the scope of this paper. We note here only that the corrections
to $F[O_2]$, $F[O_8]$ vanish as $\Lambda$ goes to zero. 
We hope to return to these issues elsewhere. But if the three-body
wave function indeed removes the low $\Lambda$ singularity, then
a relatively low value of $\Lambda$, say $0.5$ GeV, would reproduce
best the correct value of the branching ratio.

In conclusion, we have calculated the next-to-leading logarithmic 
corrections to the branching ratio of the 
exclusive $B\to K^* \gamma$ decay, using an ``inclusive'' method and
an IR cutoff parameter $\Lambda$
to remove possible infrared divergences.
For $f_B=160MeV$ and $\Lambda$=0.5 GeV, the prediction for the branching ratio
is $(5.18\pm 0.5) \cdot 10^{-5}$ and $(5.16\pm 0.5) \cdot 10^{-5}$
for two IR regularization methods we use. This corresponds to a
10\% increase over the leading order calculation. The $\Lambda$ dependence is
sizeable, around 20\%. We discuss possible ways to improve this 
uncertainty.

\vspace{0.6cm}
{\large \bf Acknowledgments}
\vspace{0.4cm} 

We thank Ch. Greub, S. Brodsky, A. Khodjamirian and R. Poghosian 
for discussions. This work was partially supported by INTAS under 
Contract INTAS-96-155 and by the Swiss Science Foundation.

\vspace{2cm}
\renewcommand{\theequation}{A.\arabic{equation}}
\setcounter{equation}0
\appendix
\begin{center}
    \bf{APPENDIX A}
\end{center}
In this Appendix we present the results for the matrix element
of inclusive $b\to s \gamma$ decay associated with the operator 
$O_7$ when we use a cutoff (method 2). The corresponding diagrams for
the wave function renormalization factors $Z_{2}^{(b)}$ and
$Z_{2}^{(s)}$  and the vertex operator are shown in Fig. 2.
Following \cite{GHW}, we introduce the renormalization scale in the 
form $\mu^2\exp(\gamma_E)/(4\pi)$. Then the $\overline{MS}$ subtraction 
corresponds
to subtracting poles in $\epsilon$ in the dimensional regularization scheme.

First we present the results for nonzero gluon mass and without cutoff. 
For that case  $Z_{2}^{(b)}$ is given by:
\begin{eqnarray}
\nonumber
&&\Sigma^{(b)}=\frac{1}{i}\frac{\alpha_s}{3\pi^3}(2\pi)^{2\epsilon}
\left (\frac{\mu^2}{4\pi}\right )^{\epsilon} \exp(\epsilon \gamma_E)
\int_{-\infty}^{\infty} d^dq \int_0^1 dx
\frac{4m_b-2(\hat{p^{'}}-\hat{q})}{(q^2-2xp^{'}\cdot q-r^2)^2}\\
&&(Z_{2}^{(b)})^{-1}-1=-\frac{d\Sigma^{(b)}}{d\hat{p^{'}}}_{\hat{p^{'}}=m_b},
\end{eqnarray}
where $\mu_0$ is the gluon mass, $r^2=xm_b^2+(1-x)\mu_0^2-xp^{'2}$.
Then the integral over q is performed and we obtain:
\begin{eqnarray}
\nonumber
&&Z_{2}^{(b)}=1+\frac{\alpha_s}{3\pi}\left (1 +\int_0^1 dx\left
(-4(1-x)\log\frac{\mu^2}{\Delta_b}
+8x(1-x^2)\frac{m_b^2}{\Delta_b}\right ) \right )\\
&&\Delta_b=x^2m_b^2+(1-x)\mu_0^2
\end{eqnarray}
The expression for $Z_{2}^{(s)}$ can be obtained from (A.2) via
the substitution $m_b \to m_s$.
The contribution of the vertex diagram Fig 2b is given by
\begin{equation}
V=\frac{2\alpha_s}{3i\pi^3}\frac{em_b}{8\pi^2}\int d^4q \int_0^1 dx_1
\int_0^{x_1} dx_2\frac{4\hat{\epsilon}\hat{k}R(p_b\cdot p_s)-
2(\hat{\epsilon}\hat{k}R\hat{q}\hat{p_s}+
\hat{p_b}\hat{q}\hat{\epsilon}\hat{k}R)}
{(q^2-2(p\cdot q)-t^2)^3}
\end{equation}
where
$p=(x_1-x_2)p_b+(1-x_1)p_s$,
$t^2=(x_1-x_2)(m_b^2-p_b^2)+(1-x_1)(m_s^2-p_s^2)+x_2\mu_0^2$.
After integrating over q, V can be expressed in the following form:
\begin{eqnarray}
&&V=<s\gamma|O_7|b>_{tree}V_1\\
\nonumber
&&V_1=-\frac{2\alpha_s}{3\pi}
\int_0^1 dx_1 \int_0^{x_1} dx_2 \frac{x_2(m_b^2+m_s^2)}{x_2\mu_0^2+
(1-x_2)((x_1-x_2)m_b^2+(1-x_1)m_s^2)}.
\end{eqnarray}
Finally, the contribution of $O_7$ for the 
case of nonzero gluon mass is given by
\begin{equation}
<s\gamma|O_7|b>_{\mu_0}=<s\gamma|O_7|b>_{tree} \left
(1+V_1+\frac{8\alpha_s}{4\pi} \log\frac{m_b}{\mu}+\left
(\sqrt{Z_{2}^{(b)}Z_{2}^{(s)}}-1
\right ) \right ).
\end{equation}
Now we  proceed to the case when we introduce a 
cutoff parameter $\Lambda$. We calculate 
the integrals for the region $|q^2|<\Lambda^2$ and use then the 
previous results to obtain the integrals over the region $|q^2|>\Lambda^2$.
We have the following expressions:
\begin{eqnarray}
\nonumber
&&\Sigma_{\Lambda}^{(b)}=\frac{1}{i}\frac{\alpha_s}{3\pi^3}(2\pi)^{2\epsilon}
\left (\frac{\mu^2}{4\pi}\right )^{\epsilon} \exp(\epsilon \gamma_E)
\int_{|q^2|<\Lambda^2}d^dq \int_0^1 dx
\frac{4m_b-2(\hat{p^{'}}-\hat{q})}{(q^2-2x(p^{'}\cdot
q)-r^2)^2}\\
&&\frac{1}{(Z_{2}^{(b)})_{|q^2|<\Lambda^2}}-1=-\frac{d\Sigma^{(b)}_{\Lambda}}
{d\hat{p^{'}}}_{\hat{p^{'}}=m_b},
\end{eqnarray}
\begin{equation}
V_{|q^2|<\Lambda^2}=\frac{2\alpha_s}{3i\pi^3}\frac{em_b}{8\pi^2}
\int_{|q^2|<\Lambda^2} d^4q \int_0^1 dx_1
\int_0^{x_1} dx_2\frac{4\hat{\epsilon}\hat{k}R(p_b\cdot p_s)-
2(\hat{\epsilon}\hat{k}R\hat{q}\hat{p_s}+
\hat{p_b}\hat{q}\hat{\epsilon}\hat{k}R)}
{(q^2-2(p\cdot q)-t^2)^3}
\end{equation}
We have the following integrals:
\begin{eqnarray}
\nonumber
&&\Sigma_{2\Lambda}^{(b1)}=\frac{2}{3\pi^3}\int_{|q^2|<\Lambda^2} d^4q 
\frac{1}{(q^2-2x(p^{'}\cdot q)-r^2)^2}\\
&&\\
\nonumber
&&\Sigma_{2\Lambda}^{(b2)}=\frac{2}{3\pi^3}\int_{|q^2|<\Lambda^2} d^4q 
\frac{q\cdot p^{'}}{p^{'2}(q^2-2x(p^{'}\cdot q)-r^2)^2}
\end{eqnarray}
\begin{eqnarray}
\nonumber
&&V_{\Lambda}^{(1)}=\frac{4}{3\pi^3}\int_{|q^2|<\Lambda^2} d^4q \frac{1}{(q^2-2(p\cdot q)-t^2)^3}\\
&&\\
\nonumber
&&V_{\Lambda}^{(2)}=\frac{4}{3\pi^3}\int_{|q^2|<\Lambda^2} d^4q \frac{q\cdot p}{p^2(q^2-2(p\cdot q)-t^2)^3}
\end{eqnarray}
The integrals over the momentum q are
performed using a Wick rotation and we obtain:
\begin{eqnarray}
\nonumber
&&\Sigma_{2\Lambda}^{(b1)}=\frac{1}{6\pi
x^2p^{'2}}\left \{\Lambda^2+r^2-\sqrt{\Lambda^4+r^4+2\Lambda^2
(r^2+2x^2p^{'2})}-\right. \\
\nonumber
&&\left. -2x^2p^{'2}\left [{\rm arctanh} \left
(1+\frac{2x^2p^{'2}}{r^2}\right )+{\rm arctanh}\left
(\frac{-2 x^2p^{'2}-r^2-\Lambda^2}
{\sqrt{\Lambda^4+r^4+2\Lambda^2(r^2+2x^2p^{'2}}}\right ) \right
] \right \}\\
\nonumber
&&\Sigma_{2\Lambda}^{(b2)}=-\frac{1}{12\pi x^3}
\left \{ \frac{1}{p^{'4}}\left (\Lambda^4
+r^2(r^2-2x^2p^{'2})
-(\Lambda^2+r^2-2x^2p^{'2})\sqrt{\Lambda^4 +r^4+2\Lambda^2
(r^2+2x^2p^{'2})}  +\right. \right. \\
\nonumber
&&\left. \left. +2r^2\Lambda^2\right )+4x^4 \left [{\rm arctanh} \left
(1+\frac{2x^2p^{'2}}{r^2}\right )+{\rm arctanh}\left
(\frac{-2 x^2p^{'2}-r^2-\Lambda^2}
{\sqrt{\Lambda^4+r^4+2\Lambda^2(r^2+2x^2p^{'2}}}\right ) \right ]
\right \}\\
&&V_{\Lambda}^{(1)}=\frac{(-\Lambda^2(2p^2+t^2)-t^4)
\sqrt{(\Lambda^2+t^2)^2+4p^2\Lambda^2}+
t^2((\Lambda^4+t^4+\Lambda^2(4p^2+2t^2))}
{3\pi p^2(p^2+t^2)((\Lambda^2+t^2)^2+4p^2\Lambda^2)}\\
\nonumber
&&V_{\Lambda}^{(2)}=-\frac{(\Lambda^2(p^2+t^2)+t^4)
\sqrt{(\Lambda^2+t^2)^2+4p^2\Lambda^2}-\Lambda^4 (p^2+t^2)-
t^2\Lambda^2(3p^2+2t^2)-t^6}
{3\pi p^4(p^2+t^2)\sqrt{(\Lambda^2+t^2)^2+4p^2\Lambda^2}}
\end{eqnarray}
where we take $t^2=x_2\mu_0^2$, $p^2=(1-x_2)((x_1-x_2)m_b^2+(1-x_1)m_s^2)$, $r^2=(1-x) \mu_0^2$.
Then  $(Z_{2}^{(b)})_{|q^2|<\Lambda^2}$ and $V_{|q^2|<\Lambda^2}$ are equal to
($p^{''}=\sqrt{p^{'2}}$)
\begin{eqnarray}
\nonumber
&&V_{|q^2|<\Lambda^2}= <s\gamma|O_7|b>_{tree}\alpha_s
\int_0^1 dx_1 \int_0^{x_1} dx_2(m_b^2+
m_s^2)\left (V_{\Lambda}^{(1)}-(1-x_2)V_{\Lambda}^{(2)}\right )\\
&&(Z_{2}^{(b)})_{|q^2|<\Lambda^2}=1+\alpha_s\int_0^1 dx
\left (\frac{d}{d\hat{p^{'}}}
\left ((4m_b-2\hat{p^{'}})\Sigma_{2\Lambda}^{(b1)}+2\hat{p^{'}}
\Sigma_{2\Lambda}^{(b2)}\right )\right )_{\hat{p^{'}}=m_b}=\\
\nonumber
&&=1+\alpha_s \int_0^1 dx
\left (2m_b\frac{d}{dp^{''}}\Sigma_{2\Lambda}^{(b1)}-2\Sigma_{2\Lambda}^{(b1)}
+2m_b\frac{d}{dp^{''}}
\Sigma_{2\Lambda}^{(b2)}+2\Sigma_{2\Lambda}^{(b2)}\right )
\end{eqnarray}
Then the expression for $(Z_{2}^{(b)})_{|q^2|>\Lambda^2}$  is given by
the difference between $Z_2^{(b)}$ and $(Z_{2}^{(b)})_{|q^2|<\Lambda^2}$
while $V_{|q^2|>\Lambda^2}$ is given by the
difference of V and $V_{|q^2|<\Lambda^2}$. For both cases, we can set 
the gluon mass equal to zero, i.e we take the limit $\mu_0\to 0$.

\newpage
\begin{center}
{\bf \large Figure Captions}
\end{center}
\vspace{0.5cm}

Figure 1: Leading order contribution, associated with operator $O_7$.

Figure 2: Order $\alpha_s$ corrections for the b and s quark wave 
function renormalization and the operator $O_7$.

Figure 3: Order $\alpha_s$ corrections to the matrix element of  
the operator $O_7$ with gluon exchange between the quark lines.

Figure 4: Order $\alpha_s$ corrections to the matrix element of 
$O_2$ with gluon exchange between the quark lines and
photon emission from external (a) and internal (b) quark lines.
The crosses indicate the possible places of photon emission.

Figure 5: Order $\alpha_s$ corrections for the 
operator $O_8$ with gluon exchange.
The possible places of photon emission are labelled with a cross.

Figure 6: The contributions of the operators $O_i$, i=1,2,3,4,5,6 with 
quark exchange.
The possible places of photon emission are labelled with a cross.

Figure 7: The branching ratio $Br(b\to K^*\gamma /10^{-5}$ as a 
function of $\log[1/\Lambda (GeV)]$ for the two methods discussed and 
$f_B$=160MeV.

Figure 8: The branching ratio $Br(b\to K^*\gamma /10^{-5}$ 
as a function  of $f_B$ for the two methods and $\Lambda =0.2GeV$

\newpage

\begin{center}
Table 1.\\

The branching ratio $BR(B\to K^* \gamma)/10^{-5}$ in LL and NLL
approximation for  $\mu$=2.5GeV, 5GeV,
10Gev, $\Lambda$ = 1.0,0.5,0.2GeV and $f_B$=160MeV.
\end{center}
\vspace{0.5cm}
\begin{tabular}{|l|c|c|c|} \hline
&$\mu =2.5GeV$&$\mu =5GeV$&$\mu =10GeV$\\ \hline
$Br^{LL}(B\to K^*\gamma)/10^{-5}$&5.91&4.65&3.70\\ \hline
$Br^{NLL}(B\to K^*\gamma)/10^{-5}$, Method 1,$\Lambda$
=1GeV&5.35&5.59&5.48\\ \hline
$Br^{NLL}(B\to K^*\gamma)/10^{-5}$, Method 1,
$\Lambda$ =0.5GeV&4.70&5.18&5.21\\ \hline
$Br^{NLL}(B\to K^*\gamma)/10^{-5}$, Method 1,
$\Lambda$ =0.2GeV&4.55&5.09&5.16\\ \hline
$Br^{NLL}(B\to K^*\gamma)/10^{-5}$, Method 2,
$\Lambda$ =1GeV&6.21&5.97&5.74\\ \hline
$Br^{NLL}(B\to K^*\gamma)/10^{-5}$, Method 2 $\Lambda$
=0.5GeV&4.66&5.16&5.20\\ \hline
$Br^{NLL}(B\to K^*\gamma)/10^{-5}$, Method 2,
$\Lambda$ =0.2GeV&3.46&4.40&4.69\\ \hline
\end{tabular}

\newpage
\begin{figure}[htb]
\epsfysize=25cm
\epsfxsize=18cm
\vspace{1cm}
\mbox{\hskip 0.2in}\epsfbox{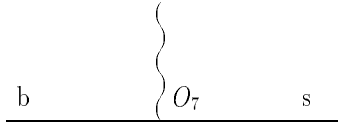}
\vspace{-23cm}
\caption{Leading order contribution, associated with operator $O_7$.}
\label{f1}
\end{figure}
\vspace{5cm}
\begin{figure}[htb]
\epsfysize=25cm
\epsfxsize=18cm
\mbox{\hskip 0.2in}\epsfbox{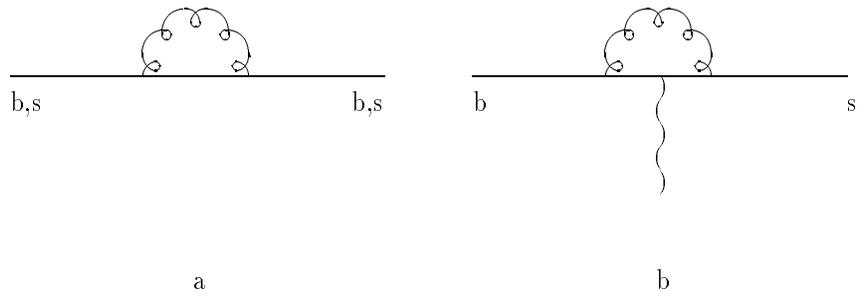}
\vspace{-19cm}
\caption{Order $\alpha_s$ corrections for the b and s quark wave 
function renormalization and the operator $O_7$.}\label{f2}
\end{figure}

\newpage

\begin{figure}[htb]
\epsfysize=25cm
\epsfxsize=18cm
\vspace{-4cm}
\mbox{\hskip 0.2in}{\vskip 0.5in}\epsfbox{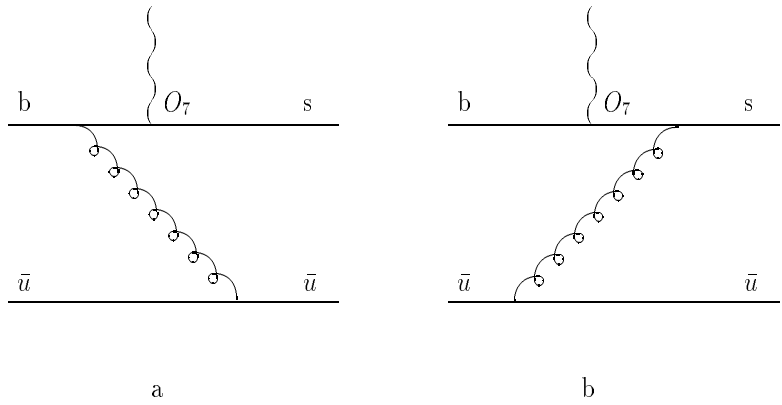}
\vspace{-17cm}
\caption{Order $\alpha_s$ corrections to the matrix element of  
the operator $O_7$ with gluon exchange between the quark lines.}
\label{f3}
\end{figure}

\vspace{3cm}
\begin{figure}[htb]
\epsfysize=25cm
\epsfxsize=18cm
\mbox{\hskip 0.2in}\epsfbox{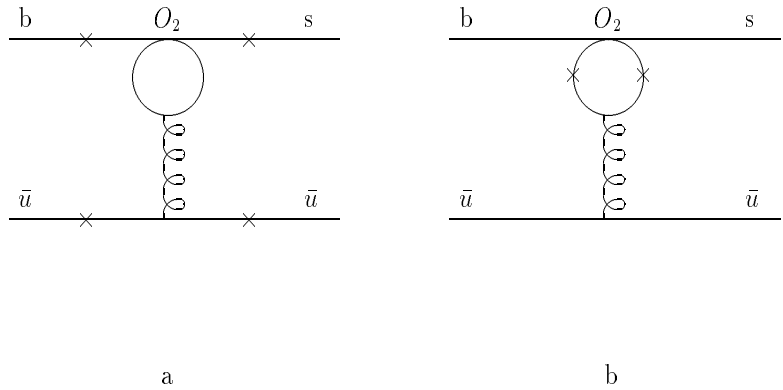}
\vspace{-19cm}
\caption{Order $\alpha_s$ corrections to the matrix element of 
$O_2$ with gluon exchange between the quark lines and
photon emission from external (a) and internal (b) quark lines.
The crosses indicate the possible places of photon emission.}\label{f4}
\end{figure}

\newpage

\begin{figure}[htb]
\epsfysize=25cm
\epsfxsize=18cm
\vspace{0cm}
\mbox{\hskip 0.2in}\epsfbox{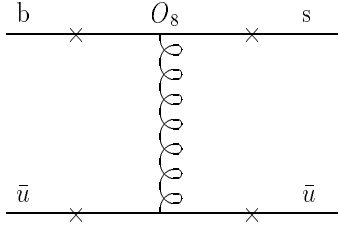}
\vspace{-20.5cm}
\caption{Order $\alpha_s$ corrections for the 
operator $O_8$ with gluon exchange.
The possible places of photon emission are labelled with a cross.}
\label{f5}
\end{figure}

\vspace{1cm}
\begin{figure}[htb]
\epsfysize=25cm
\epsfxsize=18cm
\mbox{\hskip 0.2in}\epsfbox{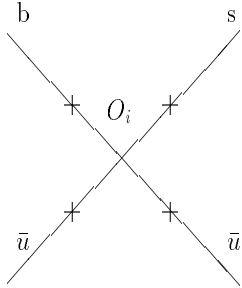}
\vspace{-17cm}
\caption{ The contributions of the operators $O_i$, i=1,2,3,4,5,6 with 
quark exchange.
The possible places of photon emission are labelled with a cross.}\label{f6}
\end{figure}

\newpage
\begin{figure}[htb]
\epsfysize=9cm
\epsfxsize=15cm
\mbox{\hskip 0.2in} {\vskip 0.5in}\epsfbox{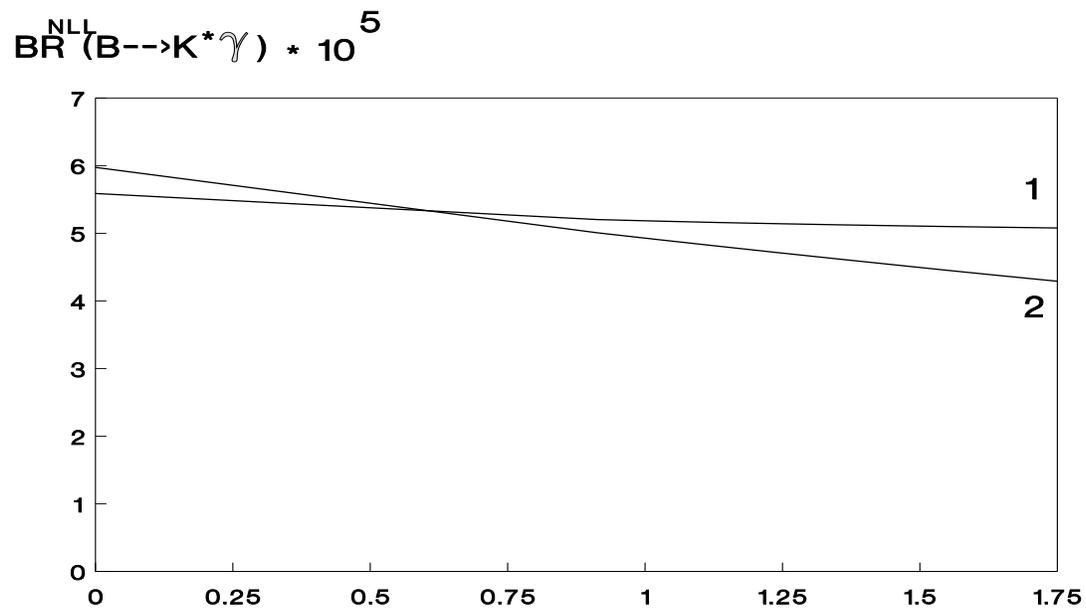}
\vspace{1cm}
\caption{The branching ratio $Br(b\to K^*\gamma /10^{-5}$ as a 
function of $\log[1/\Lambda (GeV)]$ for the two methods discussed and 
$f_B$=160MeV.}\label{f7}
\end{figure}

\newpage
\begin{figure}[htb]
\epsfysize=9cm
\epsfxsize=15cm
\mbox{\hskip 0.2in}{\vskip 0.5in}\epsfbox{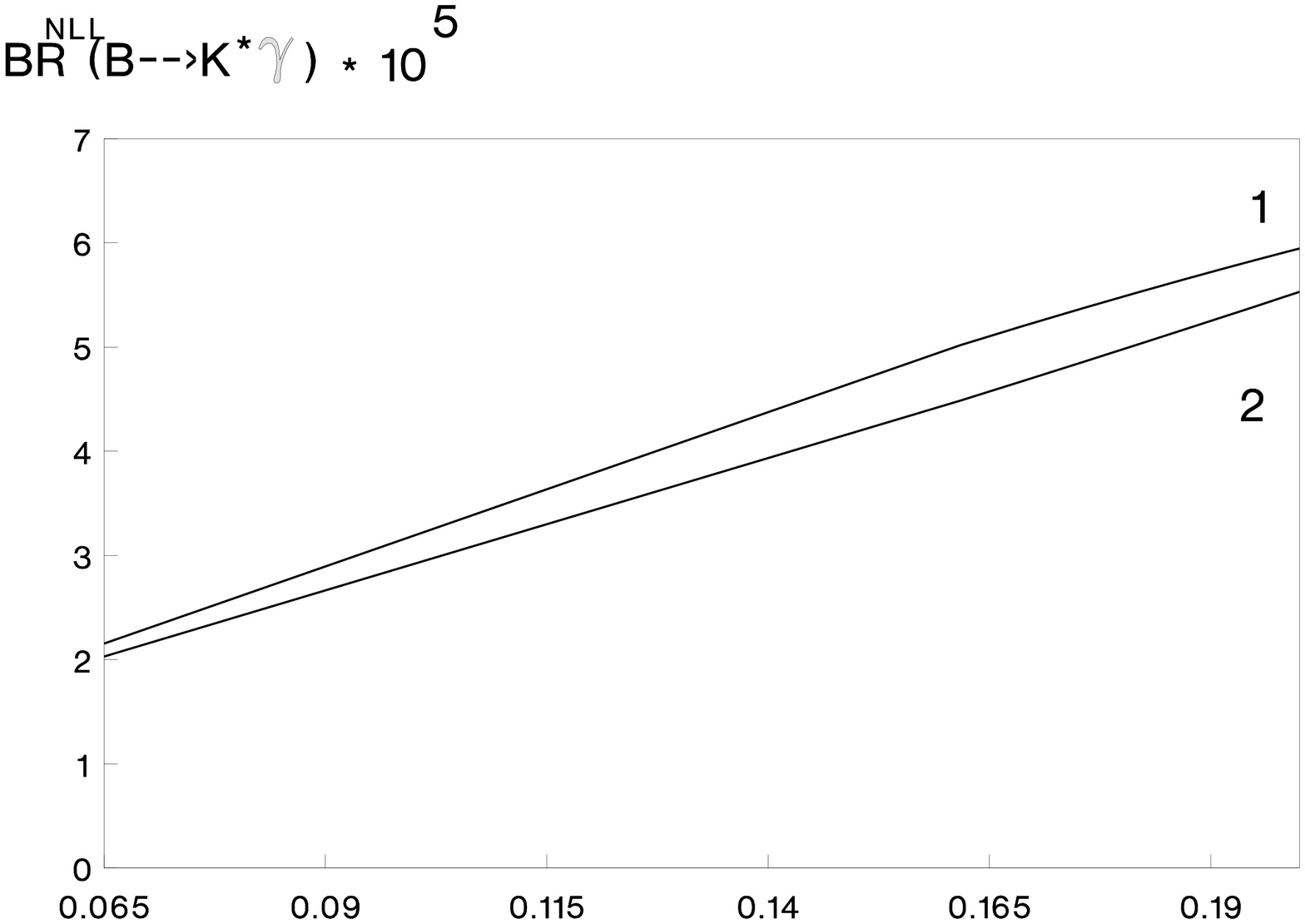}
\vspace{1cm}
\caption{The branching ratio $Br(b\to K^*\gamma /10^{-5}$ 
as a function  of $f_B$ for the two methods and $\Lambda =0.2GeV$.}
\label{f8}
\end{figure}


\vspace{2cm}
\begin{thebibliography} {99}
\bibitem{joan}
J. Hewett et al., in ``The BABAR physics book'', SLAC-Report-504,
October, 1998.
M. Neubert,  Talk given at ICHEP98, Vancouver, Canada 1998,
Hep-ph/9809343.
\bibitem{AMMARCLEO}  R.Ammar et al. (CLEO Collaboration), Phys. Rev. Lett.
71 (1993) 674.
\bibitem{GSWISE} B. Grinstein, R. Springer and M. B. Wise, Nucl. Phys.
B339 (1990).
\bibitem{BMMP} A. J. Buras, M. Misiak, M. M\"unz and S. Pokorski, Nucl. Phys.
B424 (1994) 374.
\bibitem{AY} K. Adel, Y. P. Yao, Phys. Rev. D49 (1994) 495.
\bibitem{GH} C. Greub, T. Hurth, Phys. Rev. D56 (1997) 2934.
\bibitem{AG} A. Ali, C. Greub, Phys. Lett. B361 (1995) 146.
\bibitem{GHW} C. Greub, T. Hurth, D. Wyler, Phys. Rev. D54 (1996) 3350.
\bibitem{CMM} K. Chetyrkin, M. Misiak, M. M\"unz, Phys.Lett.B200 (1997) 206.
\bibitem{BKP} A. J. Buras, A Kwiatkowski and N Pott, Phys. Lett. B414
(1997) 157.
\bibitem{CDGG} M. Ciuchini, G. Degrassi, P. Gambino and G. F. Giudice, Preprint
CERN-TH-97279 (hep-ph/9710335).
\bibitem{ALAMCLEO}  M. S. Alam et al. (CLEO Collaboration),
Phys. Rev. Lett. 74 (1995) 2885.
\bibitem{wise} Z. Ligeti, M. B. Wise, Preprint FERMILAB-Pub-99/142-T.
CALT-68-2224, hep-ph/9905277.
\bibitem{GSW}  C. Greub, H. Simma, D. Wyler, Nucl.Phys. B434 (1995) 39.
\bibitem{AAG} A. Ali, H. Asatrian, C. Greub, Phys. Lett. B429 (1998) 87.
\bibitem{BURAS} A. Buras, preprint TUM-HEP-316-98, hep-ph/9806471.  
\bibitem{GW} C. Greub, D. Wyler, Phys. Lett. B295 (1992) 293.
\bibitem{other}
T. Altomari, Phys. Rev. D37 (1988) 677;\\ 
N. Isgur and M. Wise, Phys. Rev. D42 (1990) 2388; \\
D. Wyler, Nucl.Phys. (Proc. Suppl.) 7A (1989) 358;\\
A. Ali, T. Ohl and T. Mannel,  Phys. Lett B298 (1993) 195;\\
C.A. Dominguez, N. Paver and Riazuddin, 
 Phys. Lett.  B214 (1988) 459; \\
T.M. Aliev, A.A. Ovchinnikov and V.A. Slobodeniuk;\\  
Phys. Lett. B237 (1990) 569;\\
S. Narison, 
Phys. Lett. B327 (1994) 354;\\
P. Colangelo, C.A. Dominguez, G. Narduli and N. Paver, 
Phys. Lett B317 (1994) 354;\\
P. Ball, preprint hep-ph/9308244 ( unpublished);\\ 
D.R. Burford et al. (UKQCD Collab.), Nucl. Phys. 
B447 (1995) 425.
\bibitem{ABS} A. Ali, V. Braun, H . Simma, Z.Phys C63 (1994) 437.
\bibitem{EIM} G. Eilam, A. Ioannissian, R. R. Mendel, Z.Phys. C71
(1996) 95.
\bibitem{pdg} C. Caso et al., The European Physical Journal C3 (1998) 1. 
\bibitem{brod}  S J. Brodsky, Preprint SLAC-PUB-7861,  
Invited talk at Workshop on Future Directions in Quark Nuclear Physics, 
Adelaide, Australia, 9-20 March 1998, hep-ph/9806445;\\
S J. Brodsky, Patrick Huet, Phys. Lett. B417 (1998) 145. 
\bibitem{SKW} T. Skwarnicki, Preprint HEPSY 97-03, hep-ph/9712253.
\bibitem{AOKI} S. Aoki et. al., Nucl.Phys.B Proc.Suppl. 60A (1998) 89,
hep-lat/9711043.
\bibitem{neub} M. Neubert, hep-ph/9809377.
To be published in the proceedings ICHEP 98, Vancouver, Canada, 1998.  
\bibitem{anto} F. Antonuccio, S.J. Brodsky and S.
Dalley (CERN), Phys.Lett.B 412 (1997) 104, hep-ph/9705413.      

\end{thebibliography}
\end{document}